\documentclass[conference]{IEEEtran}
\IEEEoverridecommandlockouts
\usepackage{cite}
\usepackage{amsmath,amssymb,amsfonts}
\usepackage{algorithmic}
\usepackage{graphicx}
\usepackage{textcomp}
\usepackage{xcolor}
\usepackage{float}
\usepackage{rotating}
\usepackage{ upgreek }
\usepackage{mathtools}
\usepackage{stackengine}
\usepackage{subfig}
      
\usepackage{comment}

\newsavebox\mybox
\newcommand\Includegraphics[2][]{\sbox{\mybox}{%
  \includegraphics[#1]{#2}}\abovebaseline[-.5\ht\mybox]{%
  \addstackgap{\usebox{\mybox}}}}

\def\BibTeX{{\rm B\kern-.05em{\sc i\kern-.025em b}\kern-.08em
    T\kern-.1667em\lower.7ex\hbox{E}\kern-.125emX}}
    
\begin{document}
\title{An Empirical Study on Internet Traffic Prediction Using Statistical Rolling Model}

\author{\IEEEauthorblockN{Sajal Saha, Anwar Haque, and\ *Greg Sidebottom}
\IEEEauthorblockA{\textit{Department of Computer Science} \\
\textit{University of Western Ontario, London, ON, Canada}\\
\textit{*Juniper Networks, Kanata, ON, Canada}\\
Email:\{ssaha59, ahaque32\}@uwo.ca, *gsidebot@juniper.net}
}

\maketitle

\begin{abstract}
Real-world IP network traffic is susceptible to external and internal factors such as new internet service integration, traffic migration, internet application, etc. Due to these factors, the actual internet traffic is non-linear and challenging to analyze using a statistical model for future prediction. In this paper, we investigated and evaluated the performance of different statistical prediction models for real IP network traffic; and showed a significant improvement in prediction using the rolling prediction technique. Initially, a set of best hyper-parameters for the corresponding prediction model is identified by analyzing the traffic characteristics and implementing a grid search algorithm based on the minimum Akaike Information Criterion (AIC). Then, we performed a comparative performance analysis among AutoRegressive Integrated Moving Average (ARIMA), Seasonal ARIMA (SARIMA), SARIMA with eXogenous factors (SARIMAX), and Holt-Winter for single-step prediction. The seasonality of our traffic has been explicitly modeled using SARIMA, which reduces the rolling prediction Mean Average Percentage Error (MAPE) by more than 4\% compared to ARIMA (incapable of handling the seasonality). We further improved traffic prediction using SARIMAX to learn different exogenous factors extracted from the original traffic, which yielded the best rolling prediction results with a MAPE of 6.83\%. Finally, we applied the exponential smoothing technique to handle the variability in traffic following the Holt-Winter model, which exhibited a better prediction than ARIMA (around 1.5\% less MAPE). The rolling prediction technique reduced prediction error using real Internet Service Provider (ISP) traffic data by more than 50\% compared to the standard prediction method.     
\end{abstract}


\section{Introduction}
Internet traffic engineering deals with the technology, principle, technique, and tool that assist network administrators in evaluating and optimizing the operational IP network performance. Also the traffic prediction and forecasting are some of the most crucial parts affecting network performance \cite{otoshi2015traffic}. It helps enhance the network Quality of Service (QoS) and Quality of Experience (QoE). Also, traffic forecasting assists ISP providers in their business decisions such as new product development and service decommissions, advertising, pricing, traffic migration, etc., based on traffic forecasting results. In addition, accurate forecasting helps service providers in capacity planning and investment optimization. Therefore, selecting an appropriate methodology for network traffic prediction is critical for ISP business.

Currently, ISP providers highly depend on experienced network administrators, and they follow an instinctive approach to forecast future traffic using the market analysis data such as a possible number of customers and their usage behavior \cite{papagiannaki2005long}. The factors they considered for their prediction can be divided into two main categories: internal and external factors. Internal factors are related to ISP companies, such as introducing new services, traffic migration, speed up-gradation, etc. In contrast, external factors come from outside, such as new internet applications, regional economic factors, seasonal effects, etc. Therefore, the intuitive method can only predict a rough estimation of future traffic, which is inadequate to make business decisions.

On the other hand, Operational Research, Statistics, and Computer Science contributions led to reliable prediction methods that replaced intuition-based ones. In particular, Time Series Forecasting (TSF), also termed as univariate forecasting, discusses scientific ways to predict chronologically ordered data, called time-series data \cite{shen2020novel}. The ultimate objective of TSF is to model a complex forecasting system, predicting future behavior based on the historical observation. The TSF models can be categorized into three main categories based on their learning techniques: statistical model, machine learning, and deep learning model. The prediction model from these categories requires historical data to learn the general trend in time series and make inferences about the future. The learning models also demand different settings of hyper-parameters based on the dataset size and complexity. The prediction model's learning capability and accuracy directly rely on these parameter configurations. 

Traditional statistical forecasting models such as Auto-Regressive (AR), Auto-Regressive Moving Average (ARMA), Auto-Regressive Integrated Moving Average (ARIMA), AutoRegressive Integrated Moving Average (ARIMA), Seasonal Autoregressive Integrated Moving Average (SARIMA), Seasonal Auto-Regressive Integrated Moving Average with eXogenous factors (SARIMAX), Holt-Winter, etc., were studied extensively at different time-series domains. These classical models have been used in traffic load forecasting \cite{nyaramneni2021arima}, cloud traffic prediction \cite{mehdi2022cloud}, electricity load forecasting \cite{nepal2020electricity}, and so on. However, the statistical models are best at capturing the Linear and Short Range Dependencies (SRD) in time-series data but exhibit the least performance in handling Long Range Dependencies (LRD), which results in poor time-series prediction and forecasting \cite{katris2015comparing}. In addition,  they are also incapable of learning the non-linear attribute of the time-series data. As a result, many variations of the classical forecasting model such as Fractionally Integrated Autoregressive Moving Average (FARIMA) \cite{wu2020fractional}, or hybrid models such as ARIMA-GARCH \cite{tran2015multiplicative} have been proposed to improve the statistical forecasting model's performance. This research explored several state-of-the-art statistical prediction methods such as ARIMA, SARIMA, SARIMAX, and Holt-Winter to predict internet traffic volume.  Also, we implemented a technique called rolling prediction to improve the performance of the standard prediction. In rolling prediction, the model used the validation data of the recent prediction for re-training the corresponding model. The comparative analysis among different prediction models summarizes their overall performance in traffic prediction. This research shows a significant improvement in traffic prediction using the rolling technique for our statistical prediction models. The main contributions of this work are as follows:
\begin{itemize}
    \item A comparative performance analysis between classical forecasting model in traffic prediction.
    \item Extracting new features from time series data for better prediction.
    \item Achieved significant performance improvement for the traditional model by applying the rolling prediction technique. 
\end{itemize}
This paper is organized as follows. Section \ref{literature} describes the literature review of current traffic prediction using statistical models. Section \ref{method} presents the proposed methodology introducing the rolling prediction model to the traditional model. Section \ref{result} presents the performance results of the different prediction methods and draws a comparative picture between standard prediction and rolling prediction. Finally, section \ref{conclusion} concludes our paper and sheds light on future research directions.
\begin{comment}
\begin{figure}[!htbp]
\centering
    \includegraphics[width=9cm,height = 5.2cm]{all_t.jpg}
    \caption{Original traffic pattern}
    \label{fig:All traffic}
\end{figure}
\end{comment}
\section{Literature Review}
\label{literature}
Khashei and Bijari \cite{khashei2011novel} proposed an ensemble forecasting model to improve accuracy. Their model is comprised of a statistical model called Auto-Regressive Integrated Moving Average (ARIMA) and Artificial Neural Network (ANN) model. They identified the limitation of the ANN model in handling linear data, which motivated them to apply a hybrid model based on Multi-Layer Perceptron (MLP) to process the non-linear part of the time-series data. ARIMA model handles the linear component in the time series data. A hybrid forecasting model \cite{khashei2011novel} combining the ARIMA and ANN has been developed to improve overall forecasting accuracy. Their model has been tested on three well-known data sets and indicated an improved performance for all of them. The first stage of their methodology was to generate the required data using the ARIMA model, and then this data is fed into the ANN model to predict the future. U. Kumar and V. Jain \cite{kumar2010arima} investigated the performance of the Auto-Regressive Moving Average (ARMA) and ARIMA model to predict the future in advance. They fine-tune the model parameters $p$, $q$, and $d$ by experimenting with different information criteria such as AIC (Akaike Information Criterion), HIC (Hannon–Quinn Information Criterion), BIC (Bayesian Information criterion), and FPE (Final Prediction Error). They also consider AutoCorrelation function (ACF) and Partial AutoCorrelation Function (PACF) plot information to identify the best performing model. Different model performance evaluation metric such as MAPE (Mean Absolute Percentage Error), MAE (Mean Absolute Error) and RMSE (Root Mean Squared Error) has been considered in their work. M. Dastorani et al. \cite{dastorani2016comparative} performed a comparative analysis among different statistical models such AR (Auto-Regressive), MA (Moving Average), ARMA, ARIMA, and SARIMA. They decomposed the original time series to process the random component using AR, MA, and ARMA models. A trial and error approach has been adapted in their research to find out the best-performing model, and they identified the stochastic model most appropriate for their problem. H. Liu et al. \cite{liu2012comparison} compared the performance of two different hybrid models such as ARIMA-ANN and ARIMA-Kalman. These models have been applied to process the non-stationary data, and both showed better performance. The ARIMA part of the hybrid model is used to identify the architecture of the ANN in the case of the ARIMA-ANN model. In contrast, it is used to initialize the Kalman measurement for the ARIMA-Kalman model. Huda M. A. El Hag and Sami M. Sharif \cite{el2007adjusted} identified the weakness of the ARIMA model in long-term prediction and suggested an adjustment in the ARIMA model to solve the problem. Their proposed Adjusted ARIMA (AARIMA) model can handle an extra parameter called self-similarity compared to the traditional ARIMA model. They used four different Hurst estimators to calculate the self-similarity. Their proposed model shows better hourly internet traffic prediction in comparison with the ARIMA model. 

The modification of conventional statistical models or their combination has been proposed earlier to improve the prediction in internet traffic based on the publicly available dataset. In this research, we focused on improving the performance of internet traffic prediction by adapting a new training strategy instead of modifying or combining the state-of-the-art statistical models. Our experiment results show a substantial performance improvement in traffic prediction after applying the new training strategy

\section{Methodology}
\label{method}
In this section, we first introduce the real IP traffic dataset used in our experiment in subsection \ref{dataset}. The data requires some preprocessing steps explained in subsection \ref{data_preprocessing} to clean and make it compatible with the prediction model. Then, we describe the feature extraction techniques to extricate the exogenous attributes from our original dataset in subsection \ref{feature_extraction}. Next, the rolling prediction method that significantly improves our model accuracy is introduced in the subsection \ref{rolling_prediction}. After that, we explain the mathematical background of the prediction model and the performance metric to evaluate them in subsection \ref{model} and \ref{performance_metric} respectively. Finally, we summarize the configuration of our experimental environment in subsection \ref{software}.
\subsection{Dataset}
\label{dataset}
Real internet traffic telemetry on several high speed interface have been used for this experiment. The data are collected every five minutes for a recent thirty day time period. The data contains average generated traffic per five minutes in bit per second (bps). There are 8563 data samples in our dataset consisting of 29 days complete (288 data instances per day) and last day incomplete data (211 data instance for 30th day).

\subsection{Data Preprocessing}
\label{data_preprocessing}
The dataset consists of time series data for the entire thirty days collected at every 5 minutes interval. The original dataset was collected in the JSON format, which is incompatible for processing using the prediction model. So, it was converted into CSV format first before starting the model prediction. Only the timestamp (GMT) and traffic data (bps) are taken from the JSON file, and all other information is discarded. The last day data in the dataset is incomplete by the previous eight hours (approximately) data and is removed from the original time series data for the experiment. Ultimately, a total of 29 days of data were considered for developing our prediction model. In addition, there are some missing values in the dataset filled up using the mean value. Finally, the time series value unit has been changed from bps to Gbps (gigabit per second) as the original value is large for feeding into our statistical model.
\subsection{Feature Extraction}
\label{feature_extraction}
Feature engineering is an essential part of any prediction model, whether a statistical or machine learning model. Some features have been derived from the time series data for better prediction. We divided a day into a total seven different portion as mid-night (12 am-3 pm), late-night (3 am-6 am), early morning (6 am - 9 am), morning (9 am – 12 am), afternoon (12 pm -15 pm), late afternoon (15 pm - 18 pm), evening (18 pm - 21 pm),  night (21 pm - 24 pm). We introduce another feature based on weekdays and weekends since there is a chance of high traffic on weekdays. These features are particularly provided in the SARIMAX model as it is capable of handling exogenous attributes.

\subsection{Rolling Prediction}
\label{rolling_prediction}
Our dataset is divided into train and test sets for training and evaluating the prediction model, respectively. The proposed model used a prediction function, which takes the testing data indices as a parameter for immediate prediction. We looped through our test set entries to make inferences for all test instances. After evaluating each test instance, the model has been re-trained using the true observation from the test set in rolling prediction. In contrast, the standard prediction technique estimates all predictions in the test set after training the model once. This rolling prediction technique exhibits improved performance for all prediction models.

\subsection{Time Series Forecasting Models}
\label{model}
\subsubsection{ARIMA} ARIMA model was proposed by Box and Jenkins and is also known as the Box-Jenkins methodology. This model predicts the future value based on the past values of the time series, that is, its own lagged values and the lagged forecast white noises. The time series need to be stationary before applying the ARIMA model as it performs well when there is no correlation and dependency among the predictors \cite{madan2018predicting}. Total of three parameters, such as order of AR term $(p)$, order of MA term $(q)$, and the number of differencing to make the time series stationary $(d)$ are required to design the ARIMA $(p, q, d)$ model. We can express the ARIMA model mathematically as follow \cite{box2015time}:

\begin{equation}
\Phi(L)^p\ \Delta^d y_t = \upphi(L)^q\ \Delta^d\epsilon_t
\end{equation}
\begin{equation}
\Delta^d y_t = y_t^{d-1} - y_{t-1}^{d-1}
\end{equation}

\begin{itemize}
    \item Here, $y_t$ is the time series.
    \item $p$, $d$, and $q$ are referred to as the order of AR, I and MA components of the ARIMA model.
    \item $\Delta^d$ is an operator to make the $y_t$ stationary.
    \item $\Phi(L)^p$ id the lag polynomials of order $p$, and $L$ is defined as the lag operator.
    \item $\epsilon_t$ is white noise.
\end{itemize}
%\hfill\\
\subsubsection{SARIMA}SARIMA is a generalized version of the ARIMA model which can handle the seasonality in the time series data. The SARIMA model requires additional four parameters such as seasonal autoregressive order $(P)$, seasonal moving average order $(Q)$, the seasonal difference $(D)$, and the length of the seasonality period $(S)$ to process the seasonal component in the series. As a result, six parameters are necessary to define the SARIMA model where $p$, $d$, and $q$ are the same as the ARIMA model. We can express SARIMA using the following mathematical equation \cite{otoshi2013traffic}. 

\begin{equation}
\label{eq:3}
\Phi(L)^p \Phi(L^S)^P \ \Delta^d y_t \Delta_S^D y_t= \upphi(L)^q\upphi(L^S)^Q\ \Delta^d\epsilon_t \Delta_S^D\epsilon_t
\end{equation}
\begin{equation}
\Delta^d y_t = y_t^{d-1} - y_{t-1}^{d-1}
\end{equation}

\begin{itemize}
    \item Here, $y_t$ is a time series with seasonality $S$.
    \item $P,D,\text{and } Q$ represent the similar meaning of $p,q,\text{and } d$ in the ARIMA model but rather applicable for seasonal lags. 
\end{itemize}
%\hfill\\
\subsubsection{SARIMAX} SARIMAX model capable of processing the exogenous features of the time series. The exogenous attribute calculated at time $t$ impacts the not auto-regressive time series value at time $t$. We can change the SARIMA equation \ref{eq:3} above to make it an equivalent SARIMAX equation as follows \cite{9233117}. 
\begin{equation}
\begin{aligned}
&{\Phi(L)^p \Phi(L^S)^P \ \Delta^d y_t \Delta_S^D y_t= \upphi(L)^q\upphi(L^S)^Q\ \Delta^d\epsilon_t  \Delta_S^D\epsilon_t +}\\
&{\ \ \ \ \ \ \ \ \ \ \ \ \ \ \ \ \ \ \ \ \ \ \ \ \ \ \ \ \ \ \ \ \ \ \  \ \sum_{i=1}^n\beta_ix_{t}^i}
\end{aligned}
\end{equation}

\begin{itemize}
    \item Here, $x_{t}^i$ is the exogenous attribute at time $t$ and $n$ is the total number of exogenous features. 
    \item $\beta_i$ is the coefficient for the variable $x_i$
\end{itemize}

\subsubsection{Holt-Winter}The Holt-Winters method is also known as Triple Exponential Smoothing, one of the popular algorithms designed for time-series forecasting. The first studies of Exponential Smoothing are back to the Simeon Poisson; after him in 1956, Robert Brown has introduced its forecasting application. This forecasting model is defined in three equations one for level $(l_t)$, one for trend $(b_t)$, and one for seasonality $(s_t)$ and also forecasting equation with smoothing parameters $\alpha$, $\beta$, and $\gamma$, respectively. Holt-Winters method has two variations, additive and multiplicative, that differ like the seasonal data values. We used an additive version of the Holt-Winter model for our experiment, and the equation can be defined as follow \cite{talkhi2021modeling}: 
\begin{equation}
    l_t=  \alpha(y_{t}-s_{t-m} )+(1- \alpha) (l_{t-1}+ b_{t-1})
\end{equation}
\begin{equation}
    b_t = \beta (l_t - l_{t-1}) + (1 - \beta) b_{t-1}
\end{equation}
\begin{equation}
    s_t = \gamma (y_t - l_{t-1} - b_{t-1}) + (1 - \gamma) s_{t-m}
\end{equation}
\begin{equation}
\hat{y}_{t+h \mid t}= l_t+hb_t+s_{t+h-m(k+1)}
\end{equation}

\begin{itemize}
    \item Here, $m$ represents seasonal period. 
    \item $\beta_i$ is the coefficient for the variable $x_i$
    \item $h$ is forecast horizon
    \item $k$ is the integer part of $(h-1)/m$. 
\end{itemize}

\subsection{Evaluation Metrics}
\label{performance_metric} 
We used Mean Absolute Percentage Error (MAPE), to estimate the performance of our traffic forecasting models. The performance metric identify the deviation of the predicted result from the original data. For example, MAPE error represents the average percentage of fluctuation between the actual value and predicted value. Therefore, we can define our performance metric mathematically as follow: 

\begin{equation}
    MAPE = \dfrac{1}{n}\sum_{i=1}^n \bigg| \dfrac{p_i-o_i}{o_i} \bigg| \times 100 \%
\end{equation}
Here, $p_i$ and $o_i$ are predicted and original value respectively, and $n$ is the total number of test instance

\subsection{Software and Hardware Preliminaries}
\label{software}
We have used Python and the statistical model library statsmodels \cite{seabold2010statsmodels} to conduct the experiments.  Our computer has the configuration of Intel (R) i3-8130U CPU@2.20GHz, 8GB memory, and a 64-bit Windows operating system. 
\begin{table}[!htbp]
\caption{Top five best performing model parameter}
\label{tab:Top five best performing model parameter}
\centering
\begin{tabular}{lcc} 
\hline
SL & ~ ~(p, d, q)~ ~ & ~ ~ ~AIC~ ~~  \\ 
\hline
1  & (13, 1, 16)     & 3782.588307      \\
2  & (21, 1, 2)     & 3782.751381   \\
3  & (14,1,17)      & 3783.706900   \\
4  & (13, 1, 18)     & 3785.061129   \\
5  & (21, 1, 3)     & 3785.332434   \\
6  & (22, 1, 3)     & 3786.325772      \\
7  & (16, 1, 17)     & 3786.547150   \\
8  & (17,1,17)      & 3786.761965   \\
9  & (22, 1, 4)     & 3786.964070   \\
10  & (20, 1, 2)     & 3787.326614   \\
\hline
\end{tabular}
\end{table}

\begin{figure}[!htbp]
\centering
    \includegraphics[width=9cm,height = 4cm]{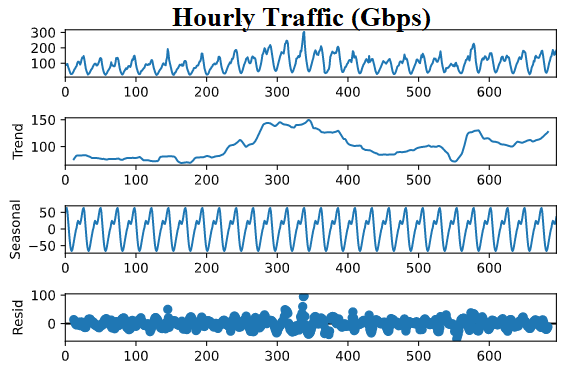}
    \caption{Decomposed time series data}
    \label{fig:decompsed_time_series}
\end{figure}

\begin{figure}
    \subfloat{\includegraphics[width=4.3cm,height=4.3cm]{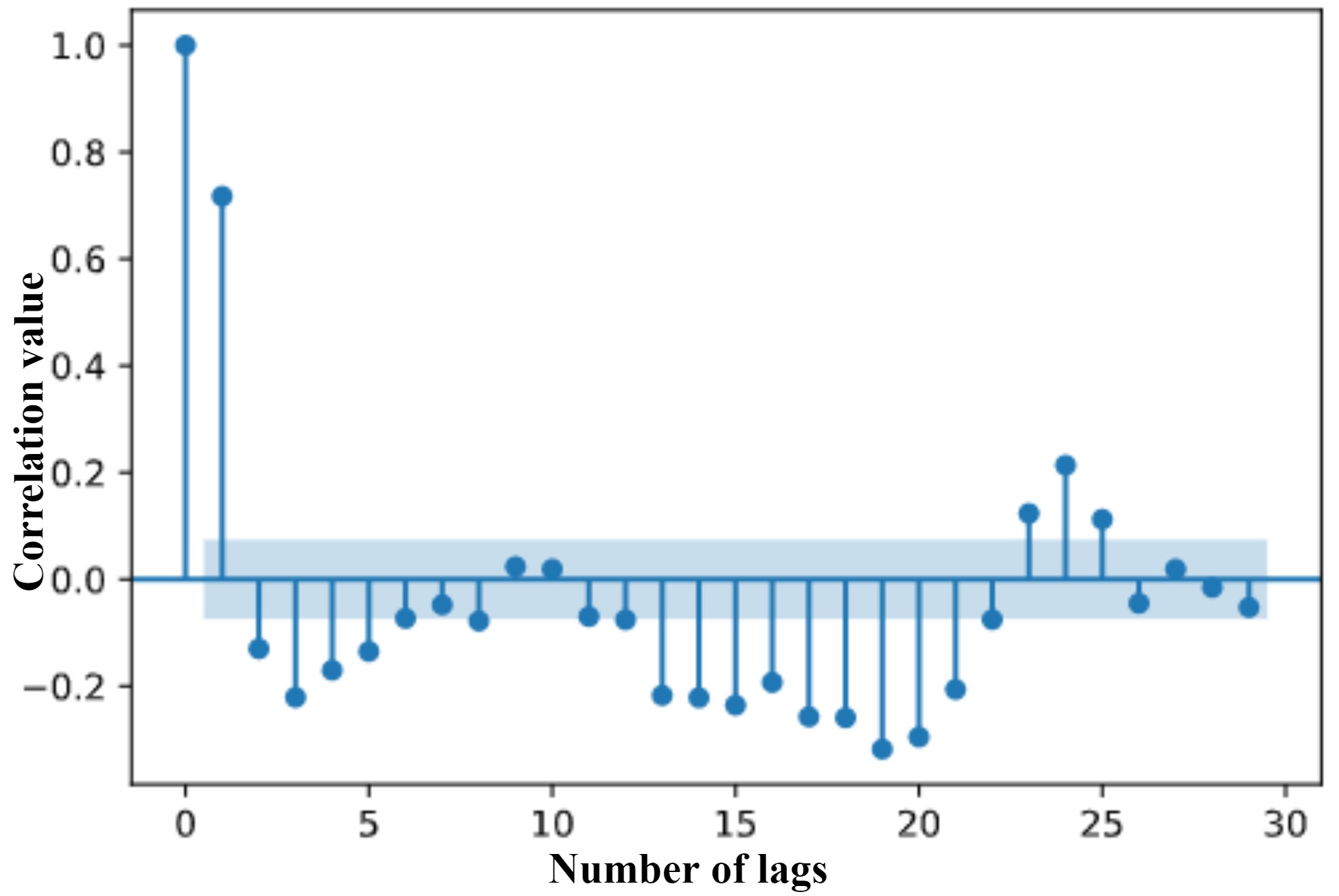}}
    \subfloat{\includegraphics[width=4.3cm,height=4.3cm]{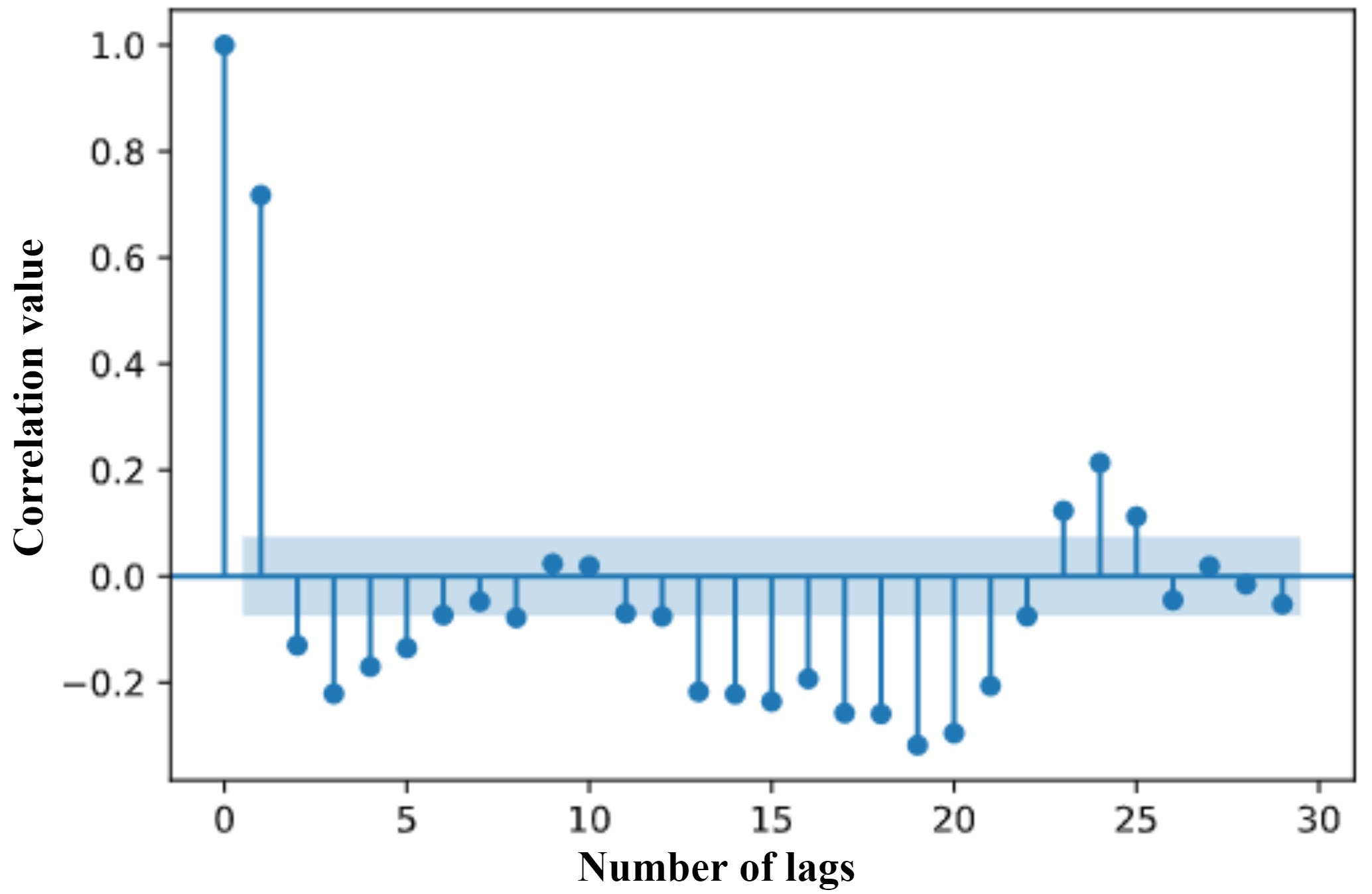}}
    \caption{Partial and Auto Partial Correlation Analysis}
    \label{fig:Partial and Auto Partial Correlation Analysis}
\end{figure}

\begin{table}
\caption{Performance summary for all model} % title of Table 
\label{tab:Performance summary for all model}
\centering      % used for centering table 
\begin{tabular}{c c c}  % centered columns (4 columns) 
\hline\hline                        %inserts double horizontal lines 
Model & MAPE (Standard Prediction) & MAPE (Rolling Prediction)\\ [0.5ex] % inserts table 
\hline                    % inserts single horizontal line 
ARIMA & 20.09\% & 7.71\%   \\    % inserting body of the table 
SARIMA &  21.03\% & 7.34\%   \\ 
SARIMAX & 21.58\% & 6.83\%    \\ 
Holt-Winter & 18.29\% & 7.59\%   \\  [1ex]       % [1ex] adds vertical space 
\hline     %inserts single line 
\end{tabular} 
\label{table:nonlin}  % is used to refer this table in the text 
\end{table} 

% \begin{table}[!htbp]
% \centering
% \caption{Performance summary for all model}
% \label{tab:Performance summary for all model}
% \begin{tabular}{llcccc} 
% \hline
% \multicolumn{4}{c}{\textbf{Forecasting}}                                         \\ 
% \hline
% \textbf{MODEL} & \textbf{~ ~ RMSE~~} & \textbf{~ ~ MAE~ ~} & \textbf{~ ~MAPE~~}  \\ 
% \hline
% BASELINE       & 18.68               & 16.37               & 16.15               \\
% ARIMA          & 24.16               & 20.58               & 17.61               \\
% SARIMA         & 25.55               & 20.63               & 14.96               \\
% SARIMAX        & 24.96               & 20.42               & 15.34               \\
% Holt-Winter    & 28.94               & 23.43               & 16.67               \\
%               &                     &                     &                     \\ 
% \hline
% \multicolumn{4}{c}{\textbf{Rolling Forecasting}}                                 \\ 
% \hline
% \textbf{MODEL} & \textbf{~ ~ RMSE~~} & \textbf{~ MAE~}     & \textbf{~ ~MAPE~~}  \\ 
% \hline
% ARIMA          & 9.3                 & 7.69                & 6.66                \\
% SARIMA         & 7.35                & 6.13                & 5.56                \\
% SARIMAX        & 7.51                & 6.08                & 5.35                \\
% Holt-Winter    & 7.24                & 5.91                & 5.24                \\
% \hline
% \end{tabular}
% \end{table}
\begin{table*}[!htbp]
\caption{Comparison between actual traffic and predicted traffic}
\label{tab:Comparison between actual traffic and predicted traffic}
\centering 
\begin{tabular}{lll}     
\hline\hline                            
Prediction Model  & ~~~~~~~~~~~~~~~~~~~~~~~~~~~~~~~~~Prediction & ~~~~~~~~~~~~~~~~~~~~~~~~~~~~~~~~~Rolling Prediction \\ [0.5ex] % inserts table heading 
\hline
        \raggedright ARIMA
        & \Includegraphics[height=1.50in,width=7.5cm]{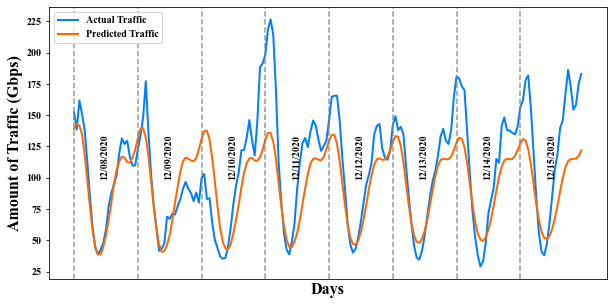}& \Includegraphics[height=1.50in,width=7.5cm]{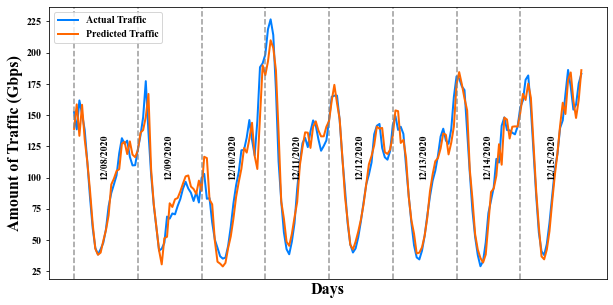} \\ 

        SARIMA 
        & \Includegraphics[height=1.50in,width=7.5cm]{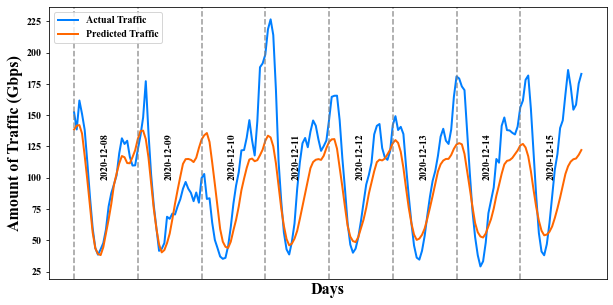}& \Includegraphics[height=1.50in,width=7.5cm]{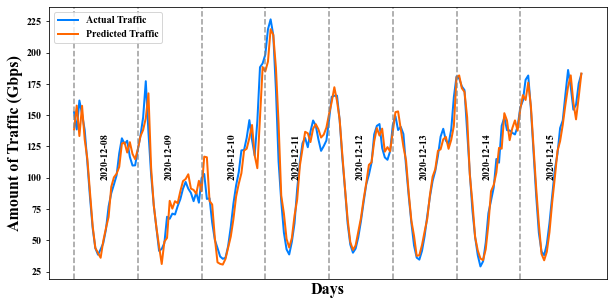} \\

        \raggedright SARIMAX
        & \Includegraphics[height=1.50in,width=7.5cm]{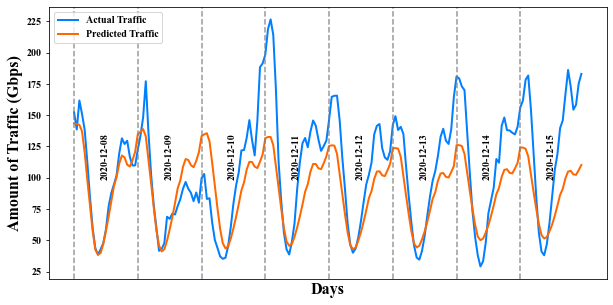}& \Includegraphics[height=1.50in,width=7.5cm]{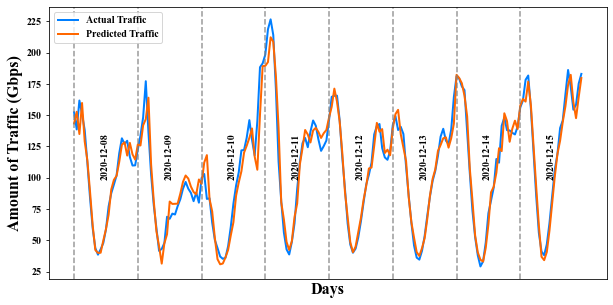} \\

        Holt-Winter
        & \Includegraphics[height=1.50in,width=7.5cm]{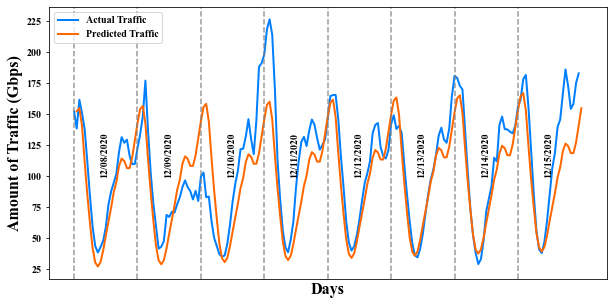}& \Includegraphics[height=1.50in,width=7.5cm]{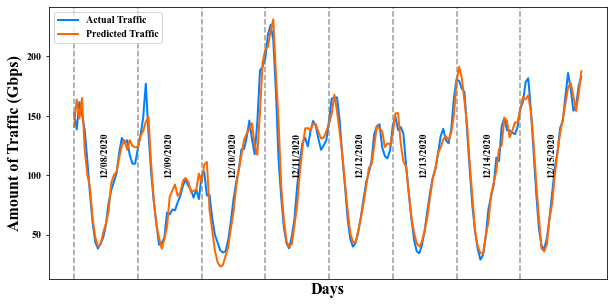} \\
\hline  
    \end{tabular}
\end{table*}

% \begin{figure}[!htbp]
% \centering
%     \includegraphics[width=9cm,height = 5cm]{images/summary.png}
%     \caption{Comparative result analysis}
%     \label{fig:Comparative result analysis}
% \end{figure}

% The total traffic pattern for 29 days is plotted in Fig. \ref{fig:All traffic}, which shows a daily pattern. 

\section{Results and Discussion}
\label{result}
We considered 21 days of data for training our model and the last eight days for testing. Before applying the prediction model, several time-series characteristics, such as stationarity, seasonality, tread, etc., need to be identified. For example, the ARIMA model performs better in stationary time-series data.  We can check the stationarity of the time series using the Augmented Dicky Fuller (ADF) test. According to the test results, the ADF statistics (-1.791022) is greater than the critical values (- 3.440), and the $p$-value (0.38) is more significant than 0.05, which satisfies its non-stationarity. To make the time series stationary, we took the first order log difference of our data and again performed the ADF test. Since the $p$-value (0.00) is less than 0.05 and ADF statistics (-19.49) is less than the critical value (-3.440), so we concluded that the time series is stationary after taking the first-order difference. This experiment helped us to set the difference order $d$ to 1 in defining our ARIMA model. Next, we decomposed our time series to identify the seasonality and trend in Fig. \ref{fig:decompsed_time_series}. There is a clear seasonality in our time series, which repeats every 24 hours that is there is a daily seasonality in our dataset. 

After that, we plotted the ACF and PACF in Fig. \ref{fig:Partial and Auto Partial Correlation Analysis} to figure out the hyper-parameters $p$ and $q$ for the ARIMA model. It is difficult to define AR($p$) and MA($q$) order from these plots as there is a sinusoidal pattern and no clear indication of the lag, which is significant for our time series data. That is why a grid search technique has been applied to identify the best hyper-parameter combination of $p$, $q$, and $d$ based on the minimum Akaike Information Criterion (AIC) value. We tested different $p$ and $q$ combinations from 0 to 24 with difference order 1 to find the best parameter set for our ARIMA model. The top ten combinations based on minimum AIC value are shown in Table \ref{tab:Top five best performing model parameter}. Our grid search result indicates the order (13, 1, 16) is the best combination with minimum AIC of 3782.588307 for the ARIMA model.  Similarly, the best model parameter we identified for SARIMA model is ($p$, $d$, $q$) = (13, 1, 16) and ($P$, $D$, $Q$, $S$) = (1, 0, 1, 24). The best performing parameter for SARIMAX model is ($p$, $d$, $q$) = (13, 1, 16) and ($P$, $D$, $Q$, $S$) = (1, 0, 1, 24). We also provided the exogenous features extracted from our original traffic data according to the feature extraction process discussed in subsection \ref{feature_extraction} into the SARIMAX model. Finally, we fine-tuned three hyper-parameters parameters for the Holt-Winter model: trend type, seasonality type, and seasonality period. The additive version of the Holt-Winter model is used in our experiment since the seasonal variation is relatively constant over time.

We experimented with two different types of prediction methodology. The standard prediction method is mainly divided into two stages: train and test. But rolling prediction model used the validation data of the latest prediction to re-train the model after each inference. The result presented in Table \ref{tab:Performance summary for all model} shows improved performance for all models in rolling prediction if we compare it with standard prediction. The model performance evaluation metrics MAPE in standard prediction decreased to more than 50\% in rolling prediction for every model. Firstly, we applied ARIMA model with the best parameter combination ($p = 13$, $d = 1$, and $q = 16$), and it resulted in average percentage prediction error of 7.71 Gbps. Since ARIMA cannot handle the seasonality in the time-series data, we implemented another model, SARIMA, accepting seasonality information as a hyper-parameter.  SARIMA provides better results with a 7.34\% average deviation from the original traffic in the rolling prediction, reducing the prediction error by more than 4\% compared to ARIMA. Next, we extracted some features from our original dataset according to the feature extraction process described in section \ref{feature_extraction}. The original traffic and their exogenous attributes are then trained using SARIMAX, which can handle extra features along with traffic seasonality. The SARIMAX yields the best prediction result in our experiment with the lowest MAPE of 6.83\%, decreasing the error by more than 11\% and 6\% compared to ARIMA and SARIMA, respectively. Finally, we implement Holt-Winter prediction model to handle the variability in our traffic data by using the exponential smoothing technique.  The Holt-Winter shows 7.59\% average fluctuation between predicted and actual traffic using rolling prediction, and it is 1.5\% less error than ARIMA. Our best-performing model SARIMAX reduces the prediction error by more than 10\% in comparison to the Holt-winter. In Table \ref{tab:Comparison between actual traffic and predicted traffic}, we depicted the actual and expected traffic for the last eight days using both standard prediction and rolling prediction. The comparison between actual and predicted traffic shows a significantly better fitting using the rolling prediction technique.

\section{Conclusion}
\label{conclusion}
In this research, we experimented with several internet traffic prediction models. We tried to improve the performance of state-of-the-art prediction models to learn the general trend in real IP traffic. Also, a comparative performance analysis among several conventional statistical models has been conducted in traffic prediction. Our experimental results show a significant improvement in traffic prediction when we feed our model with the validation data after each prediction. We considered four different prediction models, e.g., ARIMA, SARIMA, SARIMAX, and Holt-Winter, showing a better accuracy using the rolling prediction technique than the standard prediction. As part of our future work, we would like to model the residual variation using Autoregressive Conditional Heteroskedasticity (ARCH) and  Generalized AutoRegressive Conditional Heteroskedasticity (GARCH). Also, we plan to extend our work from single-step traffic prediction to multi-step prediction. 

% Finally, we will explore the machine learning and deep learning models to compare their performance and determine the best methodology for real IP traffic prediction.
\bibliographystyle{IEEEtran}
\bibliography{ref}

% Generated by IEEEtran.bst, version: 1.12 (2007/01/11)
\begin{thebibliography}{10}
\providecommand{\url}[1]{#1}
\csname url@samestyle\endcsname
\providecommand{\newblock}{\relax}
\providecommand{\bibinfo}[2]{#2}
\providecommand{\BIBentrySTDinterwordspacing}{\spaceskip=0pt\relax}
\providecommand{\BIBentryALTinterwordstretchfactor}{4}
\providecommand{\BIBentryALTinterwordspacing}{\spaceskip=\fontdimen2\font plus
\BIBentryALTinterwordstretchfactor\fontdimen3\font minus
  \fontdimen4\font\relax}
\providecommand{\BIBforeignlanguage}[2]{{%
\expandafter\ifx\csname l@#1\endcsname\relax
\typeout{** WARNING: IEEEtran.bst: No hyphenation pattern has been}%
\typeout{** loaded for the language `#1'. Using the pattern for}%
\typeout{** the default language instead.}%
\else
\language=\csname l@#1\endcsname
\fi
#2}}
\providecommand{\BIBdecl}{\relax}
\BIBdecl

\bibitem{otoshi2015traffic}
T.~Otoshi, Y.~Ohsita, M.~Murata, Y.~Takahashi, K.~Ishibashi, and K.~Shiomoto,
  ``Traffic prediction for dynamic traffic engineering,'' \emph{Computer
  Networks}, vol.~85, pp. 36--50, 2015.

\bibitem{papagiannaki2005long}
K.~Papagiannaki, N.~Taft, Z.-L. Zhang, and C.~Diot, ``Long-term forecasting of
  internet backbone traffic,'' \emph{IEEE transactions on neural networks},
  vol.~16, no.~5, pp. 1110--1124, 2005.

\bibitem{shen2020novel}
Z.~Shen, Y.~Zhang, J.~Lu, J.~Xu, and G.~Xiao, ``A novel time series forecasting
  model with deep learning,'' \emph{Neurocomputing}, vol. 396, pp. 302--313,
  2020.

\bibitem{nyaramneni2021arima}
S.~Nyaramneni, M.~A. Saifulla, and S.~M. Shareef, ``Arima for traffic load
  prediction in software defined networks,'' in \emph{Evolutionary Computing
  and Mobile Sustainable Networks}.\hskip 1em plus 0.5em minus 0.4em\relax
  Springer, 2021, pp. 815--824.

\bibitem{mehdi2022cloud}
H.~Mehdi, Z.~Pooranian, and P.~G. Vinueza~Naranjo, ``Cloud traffic prediction
  based on fuzzy arima model with low dependence on historical data,''
  \emph{Transactions on Emerging Telecommunications Technologies}, vol.~33,
  no.~3, p. e3731, 2022.

\bibitem{nepal2020electricity}
B.~Nepal, M.~Yamaha, A.~Yokoe, and T.~Yamaji, ``Electricity load forecasting
  using clustering and arima model for energy management in buildings,''
  \emph{Japan Architectural Review}, vol.~3, no.~1, pp. 62--76, 2020.

\bibitem{katris2015comparing}
C.~Katris and S.~Daskalaki, ``Comparing forecasting approaches for internet
  traffic,'' \emph{Expert Systems with Applications}, vol.~42, no.~21, pp.
  8172--8183, 2015.

\bibitem{wu2020fractional}
F.~Wu, C.~Cattani, W.~Song, and E.~Zio, ``Fractional arima with an improved
  cuckoo search optimization for the efficient short-term power load
  forecasting,'' \emph{Alexandria Engineering Journal}, vol.~59, no.~5, pp.
  3111--3118, 2020.

\bibitem{tran2015multiplicative}
Q.~T. Tran, Z.~Ma, H.~Li, L.~Hao, and Q.~K. Trinh, ``A multiplicative seasonal
  arima/garch model in evn traffic prediction,'' \emph{International Journal of
  Communications, Network and System Sciences}, vol.~8, no.~4, p.~43, 2015.

\bibitem{khashei2011novel}
M.~Khashei and M.~Bijari, ``A novel hybridization of artificial neural networks
  and arima models for time series forecasting,'' \emph{Applied Soft
  Computing}, vol.~11, no.~2, pp. 2664--2675, 2011.

\bibitem{kumar2010arima}
U.~Kumar and V.~Jain, ``Arima forecasting of ambient air pollutants (o 3, no,
  no 2 and co),'' \emph{Stochastic Environmental Research and Risk Assessment},
  vol.~24, no.~5, pp. 751--760, 2010.

\bibitem{dastorani2016comparative}
M.~Dastorani, M.~Mirzavand, M.~T. Dastorani, and S.~J. Sadatinejad,
  ``Comparative study among different time series models applied to monthly
  rainfall forecasting in semi-arid climate condition,'' \emph{Natural
  Hazards}, vol.~81, no.~3, pp. 1811--1827, 2016.

\bibitem{liu2012comparison}
H.~Liu, H.-q. Tian, and Y.-f. Li, ``Comparison of two new arima-ann and
  arima-kalman hybrid methods for wind speed prediction,'' \emph{Applied
  Energy}, vol.~98, pp. 415--424, 2012.

\bibitem{el2007adjusted}
H.~M. El~Hag and S.~M. Sharif, ``An adjusted arima model for internet
  traffic,'' in \emph{AFRICON 2007}.\hskip 1em plus 0.5em minus 0.4em\relax
  IEEE, 2007, pp. 1--6.

\bibitem{madan2018predicting}
R.~Madan and P.~S. Mangipudi, ``Predicting computer network traffic: a time
  series forecasting approach using dwt, arima and rnn,'' in \emph{2018
  Eleventh International Conference on Contemporary Computing (IC3)}.\hskip 1em
  plus 0.5em minus 0.4em\relax IEEE, 2018, pp. 1--5.

\bibitem{box2015time}
G.~E. Box, G.~M. Jenkins, G.~C. Reinsel, and G.~M. Ljung, \emph{Time series
  analysis: forecasting and control}.\hskip 1em plus 0.5em minus 0.4em\relax
  John Wiley \& Sons, 2015.

\bibitem{otoshi2013traffic}
T.~Otoshi, Y.~Ohsita, M.~Murata, Y.~Takahashi, K.~Ishibashi, and K.~Shiomoto,
  ``Traffic prediction for dynamic traffic engineering considering traffic
  variation,'' in \emph{2013 IEEE Global Communications Conference
  (GLOBECOM)}.\hskip 1em plus 0.5em minus 0.4em\relax IEEE, 2013, pp.
  1570--1576.

\bibitem{9233117}
F.~Sheng and L.~Jia, ``Short-term load forecasting based on sarimax-lstm,'' in
  \emph{2020 5th International Conference on Power and Renewable Energy
  (ICPRE)}, 2020, pp. 90--94.

\bibitem{talkhi2021modeling}
N.~Talkhi, N.~A. Fatemi, Z.~Ataei, and M.~J. Nooghabi, ``Modeling and
  forecasting number of confirmed and death caused covid-19 in iran: A
  comparison of time series forecasting methods,'' \emph{Biomedical Signal
  Processing and Control}, vol.~66, p. 102494, 2021.

\bibitem{seabold2010statsmodels}
S.~Seabold and J.~Perktold, ``statsmodels: Econometric and statistical modeling
  with python,'' in \emph{9th Python in Science Conference}, 2010.

\end{thebibliography}

\vspace{12pt}

\end{document}